\documentclass[twocolumn,showpacs,preprintnumbers,amsmath,amssymb]{revtex4}

\usepackage{graphicx}
\usepackage{dcolumn}
\usepackage{bm}

\newcommand{\As}{A\!\!\!/}

\newcommand{\ps}{p\!\!\!/}

\begin{document}
\title{Instantaneous versus non-instantaneous approach to relativistic \\
 ionization of atomic hydrogen  by electron impact  }
 \author{S. Taj }\email{souad_taj@yahoo.fr}\author{B. Manaut}\email{manaut_bouzid@yahoo.fr} \author{A. Makhoute}
\affiliation{UFR de Physique Atomique Mol\'eculaire et Optique
Appliqu\'ee,
 Facult\'e des Sciences, Universit\'e Moulay
Isma\"{\i}l, BP : 4010, Beni M'hamed, Mekn\`es, Morocco.}
\author{Y. Attaourti}
\email{attaourti@ucam.ac.ma} \affiliation{ Laboratoire de
Physique des Hautes Energies et d'Astrophysique, Facult\'e des
Sciences Semlalia, Universit\'e Cadi Ayyad Marrakech, BP : 2390,
Morocco.}


\begin{abstract}
We present a theoretical model for atomic hydrogen ionization by
electron impact in the instantaneous approximation and the more
accurate non-instantaneous approach using the methods of Quantum
Electrodynamics, for the binary coplanar and the coplanar
asymmetric geometries. All electrons are described by plane wave
functions in the coplanar binary geometry but in the asymmetric
geometry the ejected electron is described by a Sommerfeld-Maue
wave function. It is shown that the two models give the same
results in the non relativistic limit for the binary coplanar
geometry where the interactions can be treated as instantaneous.
However, this is no longer true for the relativistic case where
one has to take into account both the instantaneous interaction
and the radiation interaction. These results are obtained in the
first order of perturbation theory.
\end{abstract}

\maketitle

\section{Introduction}
The interaction of electrons with atoms is the field that most
deeply probes both the structure and  reaction dynamics of a
many-body system \cite{1}. Electron-atom collisions that ionize
the target provide a large and interesting diversity of phenomena.
The reason for this is that a three-body final state allows a wide
range of kinematic regions to be investigated. These different
kinetic regions depend sensitively on different aspects of the
description of the collision. Up to now, there has been no
calculation of differential cross-sections by a method that is
generally valid. The understanding of the ionization by electron
impact has advanced by an iterative process involving experiments
and calculations that emphasize different aspects of the reaction.
Kinematic regions have been found that are completely understood
in the sense that absolute differential cross-sections in detailed
agreement with experiment can be calculated.\\
  These form the basis of a structure probe, electron momentum spectroscopy, that
is extremely sensitive to one-electron and electron-correlation
properties of the target ground states of the residual ion in the
case of heavy atoms. Other kinematic regions require a complete
description of the collision which may be facilitated by including
the boundary condition for the three charged particles in the
final states. This point is not trivial at all because there is no
separation distance at which the Coulomb forces in the three-body
system are strictly negligible. The pioneering experiments of
Ehrhardt et al \cite{2} are of this type. Electron-impact
ionization has been studied experimentally using relativistic
electrons by Dangerfield and Spicer \cite{3}, Hoffman et all
(1979) \cite{4} and Anholt (1979) \cite{5}. The measurements with
relativistic electrons have all been of total cross-sections
typically for the $K$ and $L$ shells of heavy atoms. Theoretical
models for total ionization cross-sections have been developed by
a number of authors including Scoffield (1978) \cite{6} and
Moiseiwitsh and Stockmann (1980) \cite{7}. This field has been
reviewed by Moiseiwitsh (1980) \cite{8}. Fuss, Mitroy and Spicer
(1982) \cite{9} have developed a theoretical model for the binary
(e,2e) reaction. This (e,2e) reaction is that were the outgoing
electrons have equal energy. The theoretical model they have
developed uses the impulse and relativistic plane wave Born
approximations (RPWBA). Nakel and Whelan (1999) \cite{10} have
reviewed the experimental and theoretical developments in the
study of relativistic (e,2e). They argued that the fully
relativistic distorted wave approximation (RDWBA) proposed by
Walters et al \cite{11} was the simple possible approximation that
allows to gain understanding of the relativistic (e,2e)
processes.\\
 As for electron impact ionization for a hydrogen
atomic target, the first work that relied on the model developed
by Fuss et al \cite{9}, was presented by Attaourti et al \cite{12}
who studied the importance of the relativistic electronic dressing
during the process of laser-assisted ionization of atomic hydrogen
by electron impact. Taking a zero laser electric field, one
recovers the binary (e,2e) process in the absence of the laser
field in the instantaneous approximation where a direct non
relativistic interaction potential was used. Later on, Attaourti
et al (2005) \cite{13} developed a simple semi-relativistic model
using a Sommerfeld-Maue wave function to describe the ejected
electron and the Darwin semi-relativistic wave function to
describe the hydrogen atomic target in its ground state \cite{14}
also in the instantaneous interaction. This simple model allows to
investigate both the relativistic binary (e,2e) reaction process
(RPWBA) and the semi-relativistic (e,2e) process in the coplanar
asymmetric
geometry (SRCBA).\\
  The purpose of the present work is to go
beyond the instantaneous approximation and use the propagator
approach of QED to include both the instantaneous and radiation
contributions necessary to describe more accurately the (e,2e)
processes in the relativistic domain. Indeed, it is well known
that non relativistically, interactions are instantaneous but this
is no longer true in the relativistic case. This important feature
of relativistic collision processes is explicitly contained in the
formalism of QED. This important point is clearly explained for
example by F. Gross (1999) \cite{15}, W. Greiner (1996) \cite{16}
and many others. To what extent the contribution of both the
instantaneous and radiation parts to these (e,2e) processes affect
and modify the non polarized differential cross section is the
question we
want to address and study.\\
  The organization of this paper is as
follows : in section 2, we review the RPWBA formalism in the
instantaneous approximation and give the theoretical results for
the same model within the framework of QED. In section 3, we
review the SRCBA formalism in the instantaneous approximation and
give theoretical results using full QED calculations. In section
4, we discuss some relevant results and in section 5, we end by a
conclusion. Throughout this work, we use atomic units and the
metric tensor $g_{\mu\nu}=diag(1,-1,-1,-1)$.
\section{Binary coplanar geometry}
\subsection{Theoretical model within the frame work of the
instantaneous approximation }
\noindent The transition matrix
element for the direct channel (we do not consider exchange
effects) is given by
\begin{eqnarray}
S_{fi}=-i\int dt<\psi _{p_{f}}(x_{1})\phi _{f}(x_{2})\mid
V_{d}\mid \psi _{p_{i}}(x_{1})\phi _{i}(x_{2})>\label{1}
\end{eqnarray}
where $V_{d}=1/r_{12}-1/r_{1}$ is the direct interaction potential ($%
t_{1}=t_{2}=t\Longrightarrow x_{1}^{0}=x_{2}^{0}=x^{0}$) and in the RPWBA, $%
\psi _{p_{f}}(x_{1})$ is the wave function describing the
scattered electron

\begin{eqnarray}
\psi
_{p_{f}}(x_{1})=\frac{u(p_{f},s_{f})}{\sqrt{2E_{f}V}}e^{-ip_{f}.x_{1}}\label{2}
\end{eqnarray}
given by a free Dirac solution normalized to the volume $V$. For
the incident electron, we use
\begin{eqnarray}
\psi
_{p_{i}}(x_{1})=\frac{u(p_{i},s_{i})}{\sqrt{2E_{i}V}}e^{-ip_{i}.x_{1}}\label{3}
\end{eqnarray}
For the atomic target, $\phi _{i}(x_{2})=\phi
_{i}(t,\mathbf{r}_{2})$ is the relativistic wave function of
atomic hydrogen in its ground state. For the ejected electron, we
use again a free Dirac solution normalized to the volume $V$ and
$\phi _{f}(x_{2})$ is given by :
\begin{eqnarray}
\phi _{f}(x_{2})=\psi
_{p_{B}}(x_{2})=\frac{u(p_{B},s_{B})}{\sqrt{2E_{B}V}}
e^{-ip_{B}.x_{2}}\label{4}
\end{eqnarray}

 The unpolarized triple differential cross section (TDCS) \cite{16} is then given by :
\begin{eqnarray}
\frac{d\overline{\sigma }}{dE_{B}d\Omega _{B}d\Omega _{f}}
&=&\frac{|\mathbf{p}_f||\mathbf{p}_B|}{|\mathbf{p}_i| c^{4}}
\frac{(2E_{i}E_{f}/c^{2}-p_{i}.p_{f}+c^{2})}{\mid
\mathbf{p}_{f}\mathbf{-p}
_{i}\mid ^{4}} \nonumber\\
&&\times 4E_B\Big| \Phi _{1,1/2,1/2}(\mathbf{q}=\mathbf{\Delta
-p}_{B})\nonumber\\&&-\Phi
_{1,1/2,1/2}(\mathbf{q}=\mathbf{-p}_{B})\Big| ^{2}\label{5}
\end{eqnarray}

The functions $\Phi _{1,1/2,1/2}(\mathbf{q})$ are the Fourier
transforms of the relativistic atomic hydrogen wave functions
\cite{12}, and the quantity $\mathbf{\Delta
=p}_{i}-\mathbf{p}_{f}\ $ is the momentum transfer.
\subsection{Theoretical model with the inclusion of the photon propagator}
\noindent The transition matrix element for the direct channel  in
the non-instantaneous approximation is given by :
\begin{equation}
S_{fi}=-i\int_{-\infty }^{+\infty }dt<\phi _{f}(x)|\As(x)| \phi
_{i}(x)>\label{8}
\end{equation}
where $\As(x)=\gamma_{\mu}A^{\mu}=
A_{0}(x)\gamma_{0}-\mathbf{A(x).\gamma}$. Contravariant four
vectors are written $x^\mu=(t,\mathbf{x})$. $\phi _{f}(x)=\psi
_{p_b}(x)$ is defined in Eq. (\ref{4}), and $\phi
_{i}(t,\mathbf{x})=\psi _{t}(x)$ is the relativistic wave function
of atomic hydrogen in its ground state. The electromagnetic
potential $A^{\mu}$ due to the scattered electron is given by
\begin{equation}
A^{\mu}(x)=-4\pi\int G(x-y)J^{\mu}(y)d^4y\label{9}
\end{equation}
where  $J^{\mu}(y)$ is the four-vector current for the electron
\begin{eqnarray}
J^{\mu}(y)=\overline{\psi}_{p_f}(y)\gamma^{\mu}\psi_{p_i}(y)\label{10},\qquad
\overline{\psi}_{p_f}(y)=\psi_{p_f}^{\dagger}(y)\gamma^0
\end{eqnarray}
where $\psi _{p_i}(y)$ and $\psi _{p_f}(y)$ are defined
respectively in Eq. (\ref{3}) and Eq. (\ref{4}). Using the Fourier
decomposition of the Green's function
\begin{equation}
G(x-y)=-\frac{1}{(2\pi)^4}\int
\frac{exp{[-i\kappa(x-y)]}}{\kappa^2+i\varepsilon}d^4\kappa\label{11}
\end{equation}
the scattering matrix element becomes
\begin{eqnarray}
S_{fi}&=&\int \overline{\psi}_{p_B}(x)\gamma_{\mu}\psi_t(x)\int
\int
\frac{exp{[-i\kappa(x-y)]}}{\kappa^2+i\varepsilon}\frac{d^4\kappa}{(2\pi)^4}\nonumber\\
&\times&
\overline{\psi}_{p_f}(y)\gamma^{\mu}\psi_{p_i}(y)d^4yd^4x\label{12}
\end{eqnarray}
The $y$-integration can be performed easily yielding
\begin{equation}
\int
d^4y\exp[i(\kappa+p_f-p_i)y]=(2\pi)^4\delta[\kappa-(p_i-p_f)]\label{13}
\end{equation}
Now the $\kappa$-integration is done, and the $S_{fi}$ reads with
the usual normalization
\begin{eqnarray}
S_{fi}&=&
\frac{\overline{u}(p_f,s_f)\gamma^{\mu}u(p_i,s_i)}{\sqrt{4E_fE_iV^2}(p_i-p_f)^2}
\frac{\overline{u}(p_B,s_B)\gamma_{\mu}}{\sqrt{2E_BV}}(2\pi)^{3/2}\nonumber\\
&&\times\Phi_t(\mathbf{q})
(2\pi)\delta(E_i-E_f-E_B+|\varepsilon_b|)\label{14}
\end{eqnarray}
We note
\begin{equation}
\mathcal{F}_{\mu}=\overline{u}(p_B,s_B)\gamma_{\mu}
\end{equation}
 Using the standard procedures of QED \cite{16}], one obtains for the
 spin unpolarized triple differential cross section (TDCS)
\begin{equation}
\frac{d\overline{\sigma}}{dE_B d\Omega_f
d\Omega_B}=\frac{|\mathbf{p}_f||\mathbf{p}_B|}{|\mathbf{p}_i|
c^6}\frac{1}{2(p_i-p_f)^4}\frac{1}{2}\sum_{s_is_f}\left|\mathcal{M}_{fi}\right|^2\label{15}
 \end{equation}
 with
\begin{eqnarray}
\frac{1}{2}\sum_{s_is_f}\sum_{s_{B}}\left|\mathcal{M}_{fi}\right|^2&=&\frac{1}{2}\sum_{s_is_f}\left|\mathcal{F}_{\mu}\overline{u}(p_f,s_f)\gamma^{\mu}u(p_i,s_i)\right|^2\nonumber\\
&=&2c^2\Big\{(\mathcal{F}.p_f)(\mathcal{F}^*.p_i)+(\mathcal{F}.p_i)(\mathcal{F}^*.pf)\nonumber\\
&&-\mathcal{F}\mathcal{F}^*((p_i.p_f)-c^2)\Big\}.\label{16}
 \end{eqnarray}
 Using the spinor normalization condition
 $u^+(p,s)u(p,s')=2E\delta_{ss'}$, we find for the TDCS
\begin{eqnarray}
&&\frac{d\overline{\sigma}}{dE_B d\Omega_f
d\Omega_B}=\frac{|\mathbf{p}_f||\mathbf{p}_B|}{c^4|\mathbf{p}_i|
}\frac{8}{(p_i-p_f)^4}\nonumber\\
&&\times\left[E_f(p_B.p_i)+E_i(p_B.p_f)+E_B(2(p_i.p_f)-c^2)\right]\nonumber\\
&&\times|\Phi_{1,1/2,1/2}(\mathbf{q=\Delta-p_B})|^2.\label{17}
 \end{eqnarray}
 Note that
 \begin{eqnarray}
 \left.\begin{array}{c}
 (p_i-p_f)=(E_i-E_f)/c-(\mathbf{p_i}-\mathbf{p_f}) \\
  (p_B.p_i)=E_BE_i/c^2-\mathbf{p_B}\mathbf{p_i}\\
 (p_B.p_f)=E_BE_f/c^2-\mathbf{p_B}\mathbf{p_f}\\
 (p_i.p_f)=E_iE_f/c^2-\mathbf{p_i}\mathbf{p_f}
 \end{array}\right.\label{18}
 \end{eqnarray}
 \section{Coplanar asymmetric geometry}
  \subsection{The SRCBA theoretical model in the instantaneous approximation}
 In this section, the same computation is done, the main difference
 lying in the description of the ejected electron where now,
 a Sommerfeld-Maue wave function accurate to the order $Z/c$ in the
relativistic corrections, is used. We have $\phi
_{f}(t,\mathbf{x})=\exp (-iE_{B}t)\psi _{p_{B}}^{(-)}(\mathbf{x})$
and $\psi _{p_{B}}^{(-)}(\mathbf{x})$ is given by :
\begin{eqnarray}
&&\psi _{p_{B}}^{(-)}(\mathbf{x}) =\exp (\pi \eta _{B}/2)\Gamma
(1+i\eta _{B}) \exp
(i\mathbf{p}_{B}.\mathbf{x})\nonumber \\
 &\times&\big\{\mathsf{1}_{4}-\frac{ic}{2E_{B}}%
\mathbf{\alpha} .\mathbf{\nabla }_{(2)}\big\}\, _{1}F_{1}(-i\eta
_{B},1,-i(p_{B}x+\mathbf{p}_{B}.\mathbf{x}))\nonumber \\
&\times&\frac{%
u(p_{B},s_{B})}{\sqrt{2E_{B}V}}  \label{}
\end{eqnarray}
normalized to the volume $V$. The Sommerfeld parameter is given by
\begin{equation}
\eta _{B}=\frac{E_{B}}{c^{2}p_{B}}  \label{}
\end{equation}
 The transition matrix element for the direct channel is given by
\begin{eqnarray}
S_{fi} &=&-i\int d\mathbf{x}\frac{\overline{u}(p_{f},s_{f})}{\sqrt{2E_{f}V}}%
\gamma
_{(1)}^{0}\frac{\overline{u}(p_{B},s_{B})}{\sqrt{2E_{B}V}}\gamma
_{(2)}^{0}\Big \{\,_{1}F_{1}(i\eta _{B},\nonumber\\
&&1,i(p_{B}x+\mathbf{p}_{B}.\mathbf{x}) \mathsf{1}_{4}
-\frac{i}{2cp_{B}}(\mathbf{\alpha
.p}_{B}+p_{B}\mathbf{\alpha .}\widehat{ \mathbf{x}})\nonumber\\
&\times&_{1}F_{1}(i\eta
_{B}+1,2,i(p_{B}x+\mathbf{p}_{B}.\mathbf{x}))\Big\} \varphi
^{(\pm )}(\mathbf{x})
\exp (-i\mathbf{p}_{B}.\mathbf{x})\nonumber\\ &\times&[\exp (i%
\mathbf{\Delta .x})-1] \frac{8\pi^{2} }{\Delta ^{2}} \delta
(E_{f}+E_{B}-E_{i}-\varepsilon _{b})
\nonumber\\&\times&\frac{u(p_{i},s_{i})}{\sqrt{2E_{i}V}}\exp (\pi
\eta _{B}/2)\Gamma
(1-i\eta _{B})  \label{}%
\end{eqnarray}
The spin unpolarized TDCS is given by
\begin{eqnarray}
\frac{d\overline{\sigma} }{dE_{B}d\Omega _{B}d\Omega _{f}}&=&\frac{1}{64c^{6}\pi ^{3}}%
\frac{|\mathbf{p}_{f}||\mathbf{p}_{B}|}{|\mathbf{p}_{i}|}\frac{\exp
(\pi \eta _{B})}{\Delta ^{4}}\left| \Gamma (1-i\eta _{B})\right|
^{2}\nonumber\\&&
\left| \widetilde{S}_{fi}^{(1)}+\widetilde{S%
}_{fi}^{(2),1}+\widetilde{S}_{fi}^{(2),2}\right| ^{2}  \label{27}
\end{eqnarray}
for the description of the quantities $\widetilde{S}_{fi}^{(1)}$,
$\widetilde{S }_{fi}^{(2),1}$, $\widetilde{S}_{fi}^{(2),2}$ and
for more details see \cite{13}.
\subsection{The SRCBA theoretical model with the inclusion of the photon propagator}
The transition matrix element $S_{fi}$ for direct channel in the
non-instantaneous approximation is given by
\begin{equation}
S_{fi}=-i\int_{-\infty }^{+\infty }dt<\phi _{f}(x)|\As(x)| \phi
_{i}(x)>\label{} \end{equation}
 We replace all wave functions and $\As(x)$ in $S_{fi}$
and we get :
\begin{eqnarray}
S_{fi}&=&-i\frac{4\pi2\pi\delta(E_i-E_f-E_B+|\varepsilon_b|)}{\sqrt{8V^3E_iE_fE_B}(p_i-p_f)^2}
\nonumber\\&&
\Big(\overline{u}(p_{B},s_{B})\gamma_\mu\overline{u}(p_{f},s_{f})\gamma^\mu
u(p_i,s_i)
H_1(\mathbf{q})\nonumber\\&&+\overline{u}(p_{B},s_{B})\gamma_\mu\gamma^0(\gamma^0\frac{E_B}{c}-\ps_{B})
\overline{u}(p_{f},s_{f})\gamma^\mu
\nonumber\\&&u(p_i,s_i)H_2(\mathbf{q})+\overline{u}(p_{B},s_{B})\gamma_\mu\overline{u}(p_{f},s_{f})\gamma^\mu
\nonumber\\&&u(p_i,s_i)\widetilde{H}_2(\mathbf{q})\Big)
\end{eqnarray}
The quantities $H_1(\mathbf{q=\Delta-p_B})$,
$H_2(\mathbf{q=\Delta-p_B})$,
$\widetilde{H}_2(\mathbf{q=\Delta-p_B})$ are given in \cite{8}.
This transition matrix element contains three terms, denoted
respectively by : $S_{fi}^{(1)}$, $S_{fi}^{(2,1)}$,
$S_{fi}^{(2,2)}$.
\begin{eqnarray}
S_{fi}^{(1)}=-i[\overline{u}(p_{B},s_{B})\gamma_\mu][\overline{u}(p_{f},s_{f})\gamma^\mu
u(p_i,s_i)]H_1(\mathbf{q)}
\end{eqnarray}
\begin{eqnarray}
S_{fi}^{(2,1)}&=&-i[\overline{u}(p_{B},s_{B})\gamma_\mu\gamma^0(\gamma^0\frac{E_B}{c}-\ps_{B})]
\nonumber\\ &&[\overline{u}(p_{f},s_{f})\gamma^\mu u(p_i,s_i)]
H_2(\mathbf{q)}
\end{eqnarray}
\begin{eqnarray}
S_{fi}^{(2,2)}=-i[\overline{u}(p_{B},s_{B})\gamma_\mu][\overline{u}(p_{f},s_{f})\gamma^\mu
u(p_i,s_i)] \widetilde{H}_2(\mathbf{q)}
\end{eqnarray}
We note
\begin{eqnarray}
S_{fi}^{(1)}=-i\mathcal{A}(s_B,s_f,s_i)H_1(\mathbf{q)}
\end{eqnarray}
\begin{eqnarray}
S_{fi}^{(2,1)}=-i\mathcal{B}(s_B,s_f,s_i)H_2(\mathbf{q)}
\end{eqnarray}
\begin{eqnarray}
S_{fi}^{(2,2)}=-i\mathcal{C}(s_B,s_f,s_i)\widetilde{H}_2(\mathbf{q)}
\end{eqnarray}
 We calculate the square of $S_{fi}$. We have 9 terms
:\\ $\left| S _{fi}^{(1)}\right| ^{2}$,$ \left|
S_{fi}^{(2,2)}\right| ^{2}$, $\left| {S}_{fi}^{(2,1)}\right|
^{2}$, and $S _{fi}^{(1)\dagger }{S}_{fi}^{(2,1)}$, $S
_{fi}^{(2,1)\dagger }{S}_{fi}^{(1)}$, $S _{fi}^{(2,2)\dagger
}S_{fi}^{(2,1)}$, $S _{fi}^{(2,1)\dagger } S_{fi}^{(2,2)}$, $S
_{fi}^{(2,2)\dagger } {S}_{fi}^{(1)}$, as well as $S
_{fi}^{(1)\dagger } S_{fi}^{(2,2)}$
 .\\
The different sums over spin states are given by :
\begin{eqnarray}
&&\frac{1}{2}\sum_{s_f,s_i}\sum_{s_{B}}\mathcal{A}^\dagger(s_B,s_f,s_i)
\mathcal{C}(s_B,s_f,s_i)\nonumber\\
&&=\frac{1}{2}\sum_{s_f,s_i}\sum_{s_{B}}\mathcal{C}^\dagger(s_B,s_f,s_i)
\mathcal{A}(s_B,s_f,s_i)\nonumber\\
&&=\frac{1}{2}\sum_{s_f,s_i}\sum_{s_{B}}|\mathcal{A}(s_B,s_f,s_i)|^2\nonumber\\
 &&=\frac{1}{2}\sum_{s_f,s_i}\sum_{s_{B}}|\mathcal{C}(s_B,s_f,s_i)|^2 \nonumber\\
&&=8c^2\Big[E_f(p_B.p_i)+E_i(p_B.p_f)-E_Bc^2\Big]
\end{eqnarray}
\begin{eqnarray}
&&\frac{1}{2}\sum_{s_f,s_i}\sum_{s_{B}}|\mathcal{B}(s_B,s_f,s_i)|^2\nonumber\\
&=&8(E_B^2-c^4)\Big[E_f(p_B.p_i)+E_i(p_B.p_f)-E_Bc^2\Big]
\end{eqnarray}
\begin{eqnarray}
&&\frac{1}{2}\sum_{s_f,s_i}\sum_{s_{B}}\mathcal{A}^\dagger(s_B,s_f,s_i)
\mathcal{B}(s_B,s_f,s_i)\nonumber\\&&=
\frac{1}{2}\sum_{s_f,s_i}\sum_{s_{B}}\mathcal{C}^\dagger(s_B,s_f,s_i)
\mathcal{B}(s_B,s_f,s_i)\nonumber\\&&=
\frac{1}{2}\sum_{s_f,s_i}\sum_{s_{B}}\mathcal{B}^\dagger(s_B,s_f,s_i)
\mathcal{C}(s_B,s_f,s_i)\nonumber\\&&=
\frac{1}{2}\sum_{s_f,s_i}\sum_{s_{B}}\mathcal{B}^\dagger(s_B,s_f,s_i)
\mathcal{A}(s_B,s_f,s_i)\nonumber\\&&=2c^2\Big\{8c(p_f.p_B)(p_B.p_i)-
4c(p_f.p_i)p_B^2-\nonumber\\ &&
\frac{E_B}{c}[4E_f(p_B.p_i)+4E_i(p_B.p_f)-4E_B(p_f.p_i)]\nonumber\\&&-
\frac{4}{c}[(E_B^2-c^2p_B^2)(p_i.p_f-c^2)]\Big\}
\end{eqnarray}

The spin unpolarized TDCS in the non-instantaneous approximation
is given by
\begin{eqnarray}
\frac{d\overline{\sigma }}{dE_{B}d\Omega _{B}d\Omega
_{f}}&=&\frac{1}{2^4c^{6}\pi ^{3}}
\frac{|\mathbf{p}_{f}||\mathbf{p}_{B}|}{|\mathbf{p}_{i}|}\frac{\exp
(\pi \eta _{B})}{(p_i-p_f) ^{4}}\left| \Gamma (1-i\eta
_{B})\right| ^{2}\nonumber\\&& \left|
S_{fi}^{(1)}+S_{fi}^{(2,1)}+S_{fi}^{(2,2)}\right| ^{2} \label{}
\end{eqnarray}

 \section{Results and Discussion}
 Before beginning the discussion of the results obtained, it is
 worthwhile to recall the meaning of some abbreviations that will
 appear throughout this section. The NRPWBA stands for the non
 relativistic plane wave Born approximation where non relativistic
 plane waves are used to describe the incident, scattered and
 ejected electrons. This approximation is valid only in the non
 relativistic regime and particularly in the case of the coplanar
 symmetric geometry. The NRCBA approximation stands for the non
 relativistic Coulomb Born approximation which is valid only in
 the non relativistic regime and in the case of the asymmetric
 coplanar geometry of Ehrhardt \cite{2}. The RPWBANP stands for the
 relativistic plane wave Born approximation without propagator,
 which is valid for the case of a coplanar binary geometry and uses Dirac
 free plane wave solutions as well as a direct interaction
 potential. Its results take only into account the instantaneous
 part of the interaction potential. The RPWBAWP stands for the
 relativistic plane wave Born approximation with the inclusion of the photon
 propagator.It is valid for the case of a coplanar binary geometry, uses Dirac
 free plane wave solutions and its results take into account the instantaneous
 part of the interaction as well as the radiation part of the
 interaction. The SRCBANP stands for the semi relativistic Coulomb
 Born approximation which is valid both for the coplanar
 asymmetric geometry and the binary coplanar geometry. In this
 approximation, the incident and scattered electrons are described by free Dirac
 plane waves while the ejected electron is described by a
 Sommerfeld-Maue wave function and the atomic hydrogen target is
 described by a Darwin wave function \cite{14}. However, its uses a
 direct interaction potential in the instantaneous approximation.
  Finally, the SRCBAWP stands for the semi relativistic Coulomb
 Born approximation with the full inclusion of the photon propagator.
  It is similar to the previous approximation except that it includes both the
 instantaneous and radiation parts of the interaction.\\
  Due to this number of approximations that may be confusing, we
 will focus in our discussion on the most important features of
 our model and restrain ourselves to consider in the non
 relativistic regime, the comparison of all the TDCSs in the case
 of the binary coplanar geometry. In the case of the coplanar
 asymmetric geometry, we will compare the three relevant TDCSs, the
 SRCBAWP, the SRCBANP and the NRCBA.
 \subsection{The non relativistic regime}
 \subsubsection{The binary coplanar geometry}
 \begin{figure}[h]
\includegraphics[angle=0,width=3 in,height=4.5 in]{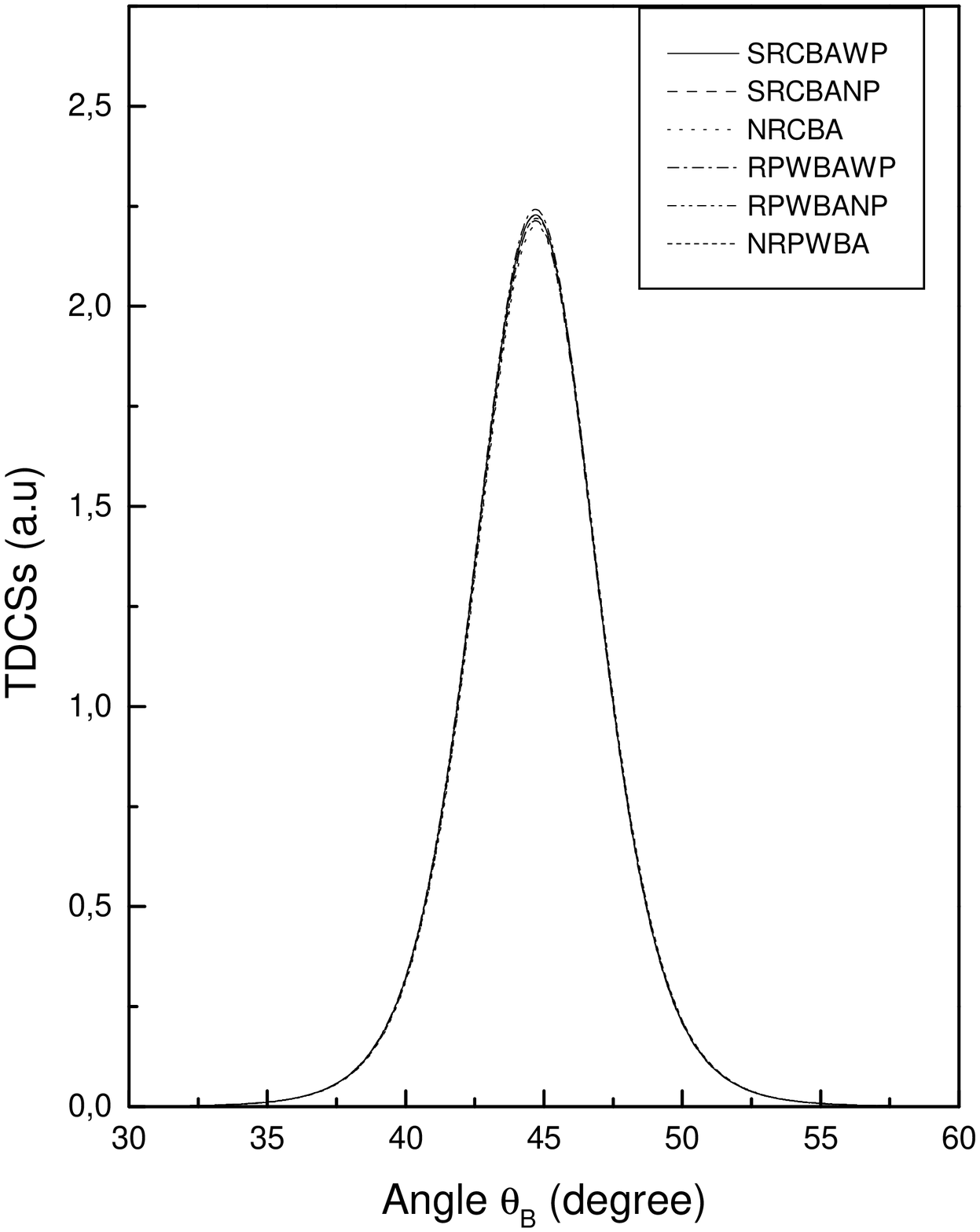}
\caption{\label{fig1.eps} TDCSs scaled in $10^{-3}$ for $E_i=2700$
$eV$ and $E_B=1349.5$ $eV$, and the angles $\theta_f=45^{\circ}$,
$\theta_i=\phi_i=\phi_f=0^{\circ}$. The angle $\phi_B$ is such
that $\phi_B=180^{\circ}$. The curves of the six approaches
overlap.}
\end{figure}
 We begin our discussion by the kinematics of the
process. For the binary coplanar geometry, we choose the following
angular situation where $p_i$ is along the $Oz$ axis
($\theta_i=\phi_i=0^{\circ}$).
 For the scattered electron, we
choose ($\theta_f=45^{\circ}, \phi_f=0^{\circ}$) and for the
ejected electron we choose $\phi_B=180^{\circ}$) and the angle
$\theta_B$
 varies from $30^{\circ}$ to $60^{\circ}$. The energies of the
 incident and ejected electrons are respectively $E_i=2700$
$eV$ and $E_B=1349.5$ $eV$. It is expected that for such a regime
and choice of geometry, relativistic effects will be small and it
is indeed the case as this can be seen in Fig.1 where the six
approaches give nearly the same results. Even if this is not
apparent on this figure, we have plotted the results of six
approaches and obtained  close results. In the binary coplanar
geometry, the TDCSs are well peaked around the specific angle
$\theta_f$ chosen and decrease very rapidly to zero within a small
angular spread (typically ten degrees). The energy of the incident
electron being non relativistic (it corresponds to a relativistic
parameter $\gamma_i=(1-(\beta_i)^2)^{(-1/2)}$), the description of
the interaction potential by a direct non relativistic potential
or the use of the photon propagator is not important in that case.
\subsubsection{The coplanar asymmetric geometry}
In this geometry, the use of any approach involving plane waves
will lead to irrelevant results since the Coulomb description of
the ejected electron is needed. The angular choice is as follows :
$p_i$ is along the $Oz$ axis and $\theta_i=0^{\circ},
\phi_i=0^{\circ}$. For the scattered electron, we choose
($\theta_f=3^{\circ},\phi_f=0^{\circ}$), for the ejected electron
we choose $\phi_B=180^{\circ}$ and the angle $\theta_B$
 varies from $-180^{\circ}$ to $180^{\circ}$.
 \begin{figure}[h]
\includegraphics[angle=0,width=3 in,height=4.5 in]{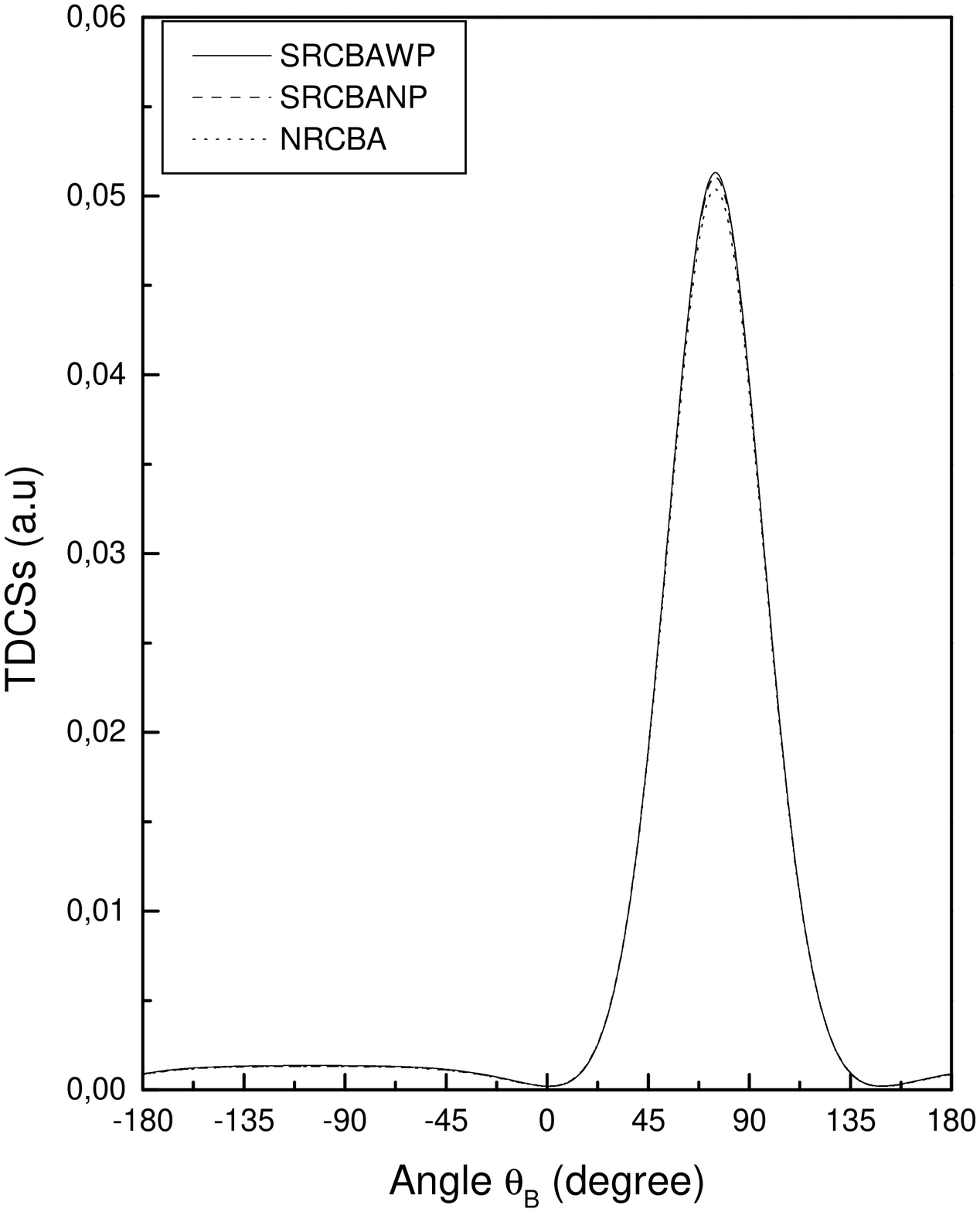}
\caption{\label{fig2.eps} TDCS for $E_i=2700$ $eV$ and $E_B=54$
$eV$, and the angles $\theta_f=3^{\circ}$,
$\theta_i=\phi_i=\phi_f=0^{\circ}$. The angle $\phi_B$ is such
that $\phi_B=180^{\circ}$. The curves overlap.}
\end{figure} The
 energy of the incident electron is $E_i=2700$ $eV$ and the energy of the
 ejected electron is $E_B=54.$ $eV$. The range of the
 angular variable $\theta_B$ is due to the fact that in the coplanar
 asymmetric geometry, one expects to see two peaks, a binary peak
 which is, in our case, located in the vicinity of $\theta_B=77^{\circ}$ and a
 recoil peak which is located in the vicinity of
 $\theta_B=-105^{\circ}$. As a validity check of our calculations,
 we have also reproduced the results of Byron and Joachain \cite{17} for the
 energies $E_i=250$ $eV$ and $E_B=5.$ $eV$. The three
 approaches : NRCBA, SRCBAWP and SRCBANP are shown in Fig. 2
 where the spread of the three TDCSs covers the whole range that
 the angular variable $\theta_B$ can take and small differences
 begin to appear particularly at the recoil peak. However and as
 might be expected for such energies, the SRCBAWP and the SRCBANP
 remain very close because for such a regime the use of the
 instantaneous or the non instantaneous approximation amounts to
 nearly the same. Radiation effects are not important enough for
 such non relativistic energies.
 \subsection{The relativistic regime}
 \subsubsection{The binary coplanar geometry}
 For this regime, the energy of the incident electron is $E_i=511002$ $eV$
 and the energy of the ejected electron is $E_B=225501$ $eV$.
 This value of the energy $E_i$ corresponds to a relativistic
 parameter $\gamma_i=2$. We choose the following
angular situation where $p_i$ is along the $Oz$ axis
($\theta_i=\phi_i=0^{\circ}$).
\begin{figure}[h]
\includegraphics[angle=0,width=3 in,height=4.5 in]{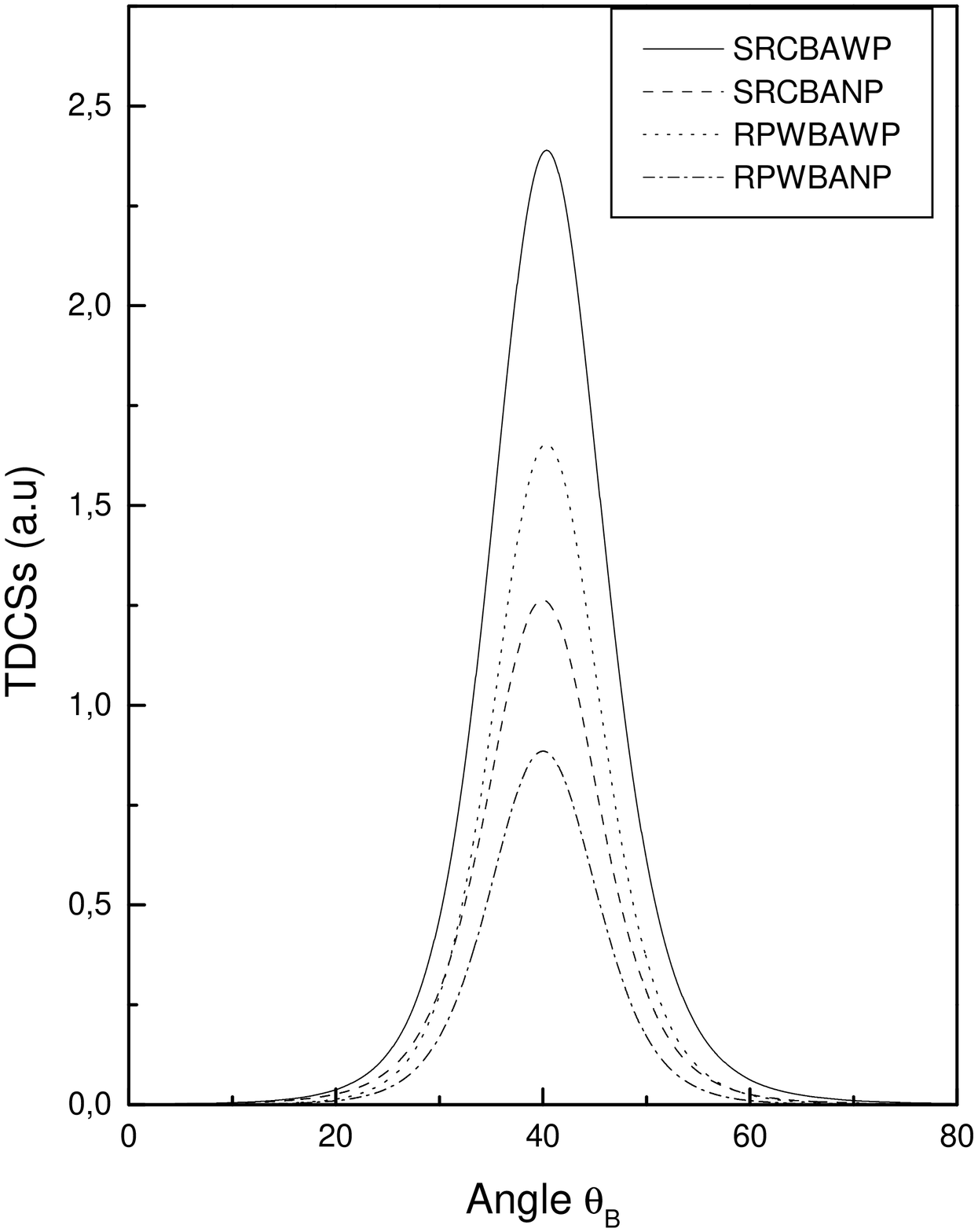}
\caption{\label{fig3.eps} TDCS scaled in $10^{-19}$ for
$E_i=510999$ $eV$ and $E_B=255499.5$ $eV$, and the angles
$\theta_f=55^{\circ}$, $\theta_i=\phi_i=\phi_f=0^{\circ}$. The
angle $\phi_B$ is such that $\phi_B=180^{\circ}$.}
\end{figure}
For the scattered electron, we choose ($\theta_f=55^{\circ}$,
$\phi_f=0^{\circ}$) and for the ejected electron we choose
$\phi_B=180^{\circ}$ and the angle $\theta_B$
 varies from $30^{\circ}$ to $60^{\circ}$.
 Relativistic effects can no longer be
 neglected and also the use of the instantaneous approximation
 becomes itself questionable. Let us explain what is contained in
 Fig 3. The first important point that has to be mentioned is the
 following : in such a regime, the use of non relativistic
 approaches is not physically founded. So, we will focus mainly on
 the four relativistic approximations aforementioned. What appears
 clearly is that the four relativistic models give different
 results with the importance of radiation effects clearly shown.
 To summarize Fig 3, we can say that as regards to the SRCBAWP and
 SRCBANP, the instantaneous approximations is no longer valid. We
 have a situation where the TDCS(SRCBAWP) is higher than the
 TDCS(SRCBANP) and also where the TDCS(RPWBAWP) is higher than the
 TDCS(RPWBANP). This is a general rule and we have made
 many simulations to assert that it is a valid rule in the
 relativistic regime.
 However, there is no general rule as to the comparison between
 the TDCS(SRCBAWP) and the TDCS(RPWBAWP) as well as the comparison
 between the TDCS(RPWBANP) and the TDCS(SRCBANP)
 because another
 choice of ($\theta_f$, e.g $\theta_f=39^{\circ}$ will give rise
 to a reverse situation where $TDCS(RPWBAWP)>TDCS(SRCBAWP)$
 whereas $TDCS(RPWBANP)>TDCS(SRCBAWP)$.

\begin{figure}[h]
\includegraphics[angle=0,width=3 in,height=4.5 in]{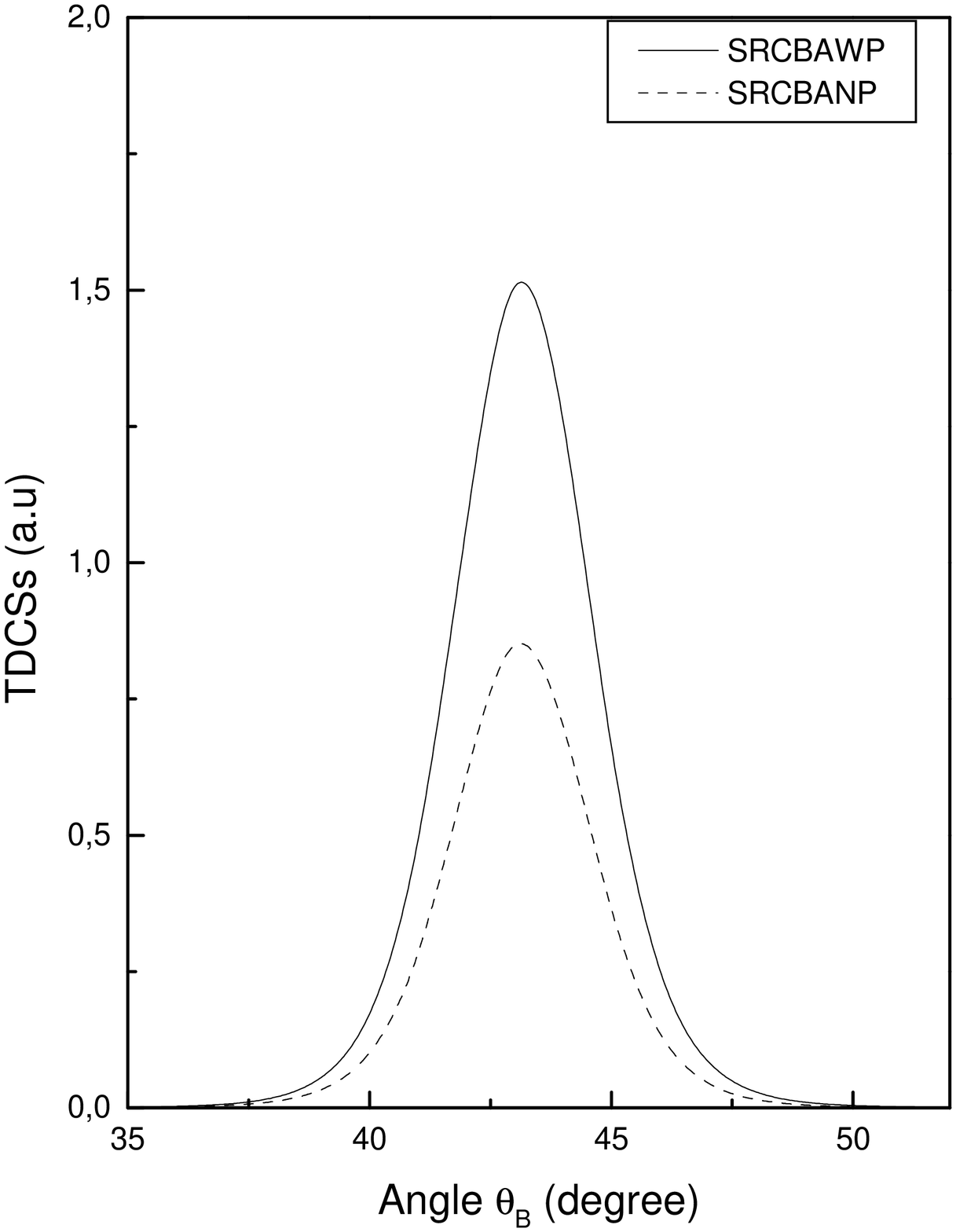}
\caption{\label{fig3.eps} TDCS scaled in $10^{-13}$ for the
energies $E_i=511002$ $eV$ and $E_B=225501$ $eV$, and the angles
$\theta_f=39^{\circ}$, $\theta_i=\phi_i=\phi_f=0^{\circ}$. The
angle $\phi_B$ is such that $\phi_B=180^{\circ}$.}
\end{figure}
 In Fig. 4, we only compare the two SRCBAWP and SRCBANP to assess
 the importance of radiation effects and as this behavior never
 changes, one can easily see that
 the use of the instantaneous approximation underestimates the
 value of the TDCS using the photon propagator by a factor two.
 The same behavior is observed with the RPWBAWP and the RPWBANP.
\subsubsection{The coplanar asymmetric geometry}
In this geometry, the angular choice is as follows : $p_i$ is
along the $Oz$ axis and $\theta_i=3^{\circ}, \phi_i=0^{\circ}$.
For the scattered electron, we choose ($\theta_f=3^{\circ},
\phi_f=0^{\circ}$), for the ejected electron we choose
$\phi_B=180^{\circ}$ and the angle $\theta_B$
 varies from $0^{\circ}$ to $180^{\circ}$. The
 energy of the incident electron is $E_i=511002$ $eV$ and the energy of the
 ejected electron is $E_B=10220.04$ $eV$.
 \begin{figure}[h]
\includegraphics[angle=0,width=3 in,height=4.5 in]{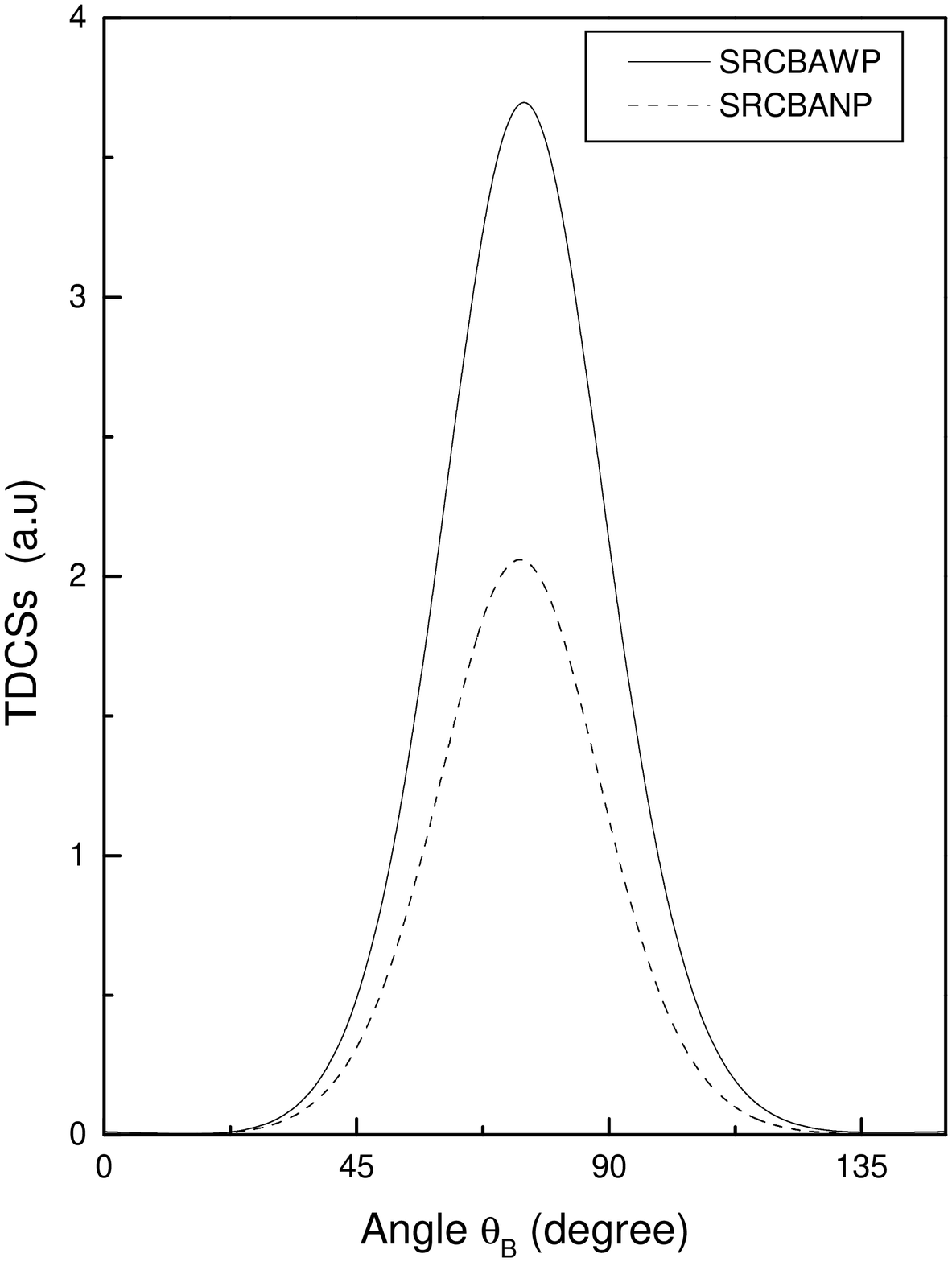}
\caption{\label{fig3.eps} TDCS scaled in $10^{-12}$ for
$E_i=511002$ $eV$ and $E_B=10220.04$ $eV$, and the angles
$\theta_f=3^{\circ}$, $\theta_i=\phi_i=\phi_f=0^{\circ}$.The angle
$\phi_B$ is such that $\phi_B=180^{\circ}$.}
\end{figure}
 In the relativistic regime, there is no occurrence of a recoil
 peak and the spread of the angular variable $\theta_B$ of the
 ejected electron is reduced. As expected, radiation effects are
 indeed important and in Fig. 5, we see that the binary peak is
 located in the vicinity of $\theta_B=75^{\circ}$ whereas the TDCS(SRCBANP) underestimates
 the  TDCS(SRCBAWP) by a factor 2. The comparison with the two
 other approaches using relativistic plane waves is not relevant
 for this geometry.
\section{Conclusion}
In this work, we have studied the two theoretical models
(instantaneous and non-instantaneous approximations) for the
relativistic ionization of atomic hydrogen by electron impact
using non relativistic or relativistic plane wave functions  to
describe all electrons in the binary coplanar geometry and using
the Sommerfed-Maue wave function to describe the ejected electron
in the case of the coplanar asymmetric geometry. In all the
approaches using relativistic plane waves, the use of the exact
relativistic wave function for atomic hydrogen in its ground state
is tractable, whereas in the asymmetric case, one can only use the
semi relativistic Darwin wave function \cite{14} to describe the
atomic target. The general conclusions that can be drawn from this
work are summarized in what follows. In the non relativistic
regime, the results obtained in the binary coplanar geometry give
the same results for an appropriate choice of the angular
parameters whereas in the coplanar asymmetric geometry, on can
only compare the three approaches NRCBA, SRCBAWP and SRCBANP. This
comparison has been made and it was shown that though relatively
small, the presence of a recoil peak together with a visible
binary peak even for an incident electron energy of $2700$ $eV$
reproduces the
qualitative features of the Ehrhardt geometry \cite{2}.\\
In the relativistic regime radiation effects are important and the
general rule that can be firmly asserted is the following : the
instantaneous approach is always smaller than the non
instantaneous approach because radiation effects can no longer be
ignored. The use of the photon propagator as well as the QED
formalism becomes necessary.


\begin{thebibliography}{90}
\bibitem{1} I.E. McCarthy and E. Weigold, \textit{ Electron-Atom
Collisions}, Cambridge University Press, 1995.
\bibitem{2} H. Ehrhardt, M. Schulz, T. Tekaat, and K. Willmann, Phys.
Rev. Lett. \textbf{22}, 89 (1969).
\bibitem{3} G.R. Dangerfield and B.M. Spicer, J. Phys. B, \textbf{8}, 1744,
(1975).
\bibitem{4} D.H.H. Hoffmann, C. Brendel, H. Genz, W. Löw, S;
Muller and A. Richter, Z. Phys., \textbf{A 293}, (1979).
\bibitem{5} R. Anholt, Phys. Rev. \textbf{A 19}, 1004 (1979).
\bibitem{6} J.H. Scofield, Phys. Rev. \textbf{A 18}, 963, (1978).
\bibitem{7} B.L. Moiseiwitsch and S.G. Stockmann, J. Phys. B, \textbf{13},
2975, (1980).
\bibitem{8} B.L. Moiseiwitsch, Adv. At. Mol. Phys., \textbf{16}, 281,
(1980).
\bibitem{9} I. Fuss, J. Mitroy, and B. M.
Spicer, J. Phys. B \textbf{15}, 3321 (1982). (1999).
\bibitem{10} W. Nakel and C.T. Whelan, Phys. Rep., 315, 409,
(1999). \textbf{315}, 409, (1999).
\bibitem{11} H.R.J. Walters, H. Ast, C.T. Whelan, R.M. Dreizler,
H. Graf, C.D. Schr\"{o}ter, J. Bonfert, W. Nakel, Z. Phys.,
\textbf{D 23}, 353, (1992).
\bibitem{12} Y. Attaourti and S. Taj, Phys. Rev. A
\textbf{69}, 063411 (2004).
\bibitem{13} Y. Attaourti, S. Taj and B. Manaut, Phys. Rev. A
\textbf{71}, 062705 (2005).
\bibitem{14} J. Eichler and W.E. Meyerhof, \textit{Relativistic Atomic
Collisions}, Academic Press, 1995.
\bibitem{15} F. Gross,
\textit{Relativistic Quantum Mechanics and Field Theory}, Wiley
Science Paperback Series, 1999.
\bibitem{16} W. Greiner, \textit{Quantum Electrodynamics}
, Second Corrected Edition, Springer, 1996.
\bibitem{17} F.W. Byron Jr and C.J. Joachain, Phys. Rep. \textbf{179}, 211,
(1989).
\end{thebibliography}
\end{document}